\documentclass[twocolumn,showpacs,showkeys,prc,aps]{revtex4}
\usepackage{graphicx}
\begin{document}
\title{Strange Disoriented Chiral Condensate} 
\author{Sean Gavin}
\affiliation{Physics Department, Wayne State University, 
Detroit, Michigan, U.S.A.}
\date{\today}
\begin{abstract}
Disoriented chiral condensate can produce novel fluctuations of
kaons as well as pions. Robust statistical observables can be used to
extract the novel fluctuations from background contributions in
$K_S^0K^\pm$ measurements in nuclear collisions. To illustrate how
this can be done, I present new event-generator computations of these
observables.
\end{abstract}
\maketitle

\section{Introduction}
\label{intro}

Relativistic nuclear collisions can produce matter in which chiral
symmetry is restored. One possible consequence of the restoration and
the subsequent re-breaking of chiral symmetry is the formation of
disoriented chiral condensates (DCC) -- transient regions in which the
average chiral order parameter differs from its value in the
surrounding vacuum \cite{DCCreview,qmrev1,qmrev2}. Previous efforts to
describe DCC signals have focused on pion production.

In Ref.~\cite{GavinKapusta}, Kapusta and I have explored the possible
influence of DCC on kaon production, inspired by the explanation due
to Kapusta and Wong \cite{KapustaWong} of measurements of $\Omega$ and
$\overline\Omega$ baryon enhancement \cite{SPSbaryons} at $17$~$A$~GeV
at the CERN SPS in terms of the production of many small DCC regions
within individual collision events. If true, Ref.~\cite{KapustaWong}
implies that the evolution of the condensate can have a significant
effect on {\it strange} particle production. The importance of strange
degrees of freedom in describing chiral restoration has been long
appreciated \cite{PisarskiWilczek,Columbia,GGP2,Karsch,Lenaghan}, but
simulations of the three flavor linear sigma model had suggested that
strange kaon fields are much less important than the pion
fields~\cite{Randrup}. Nevertheless, the $\Omega$ and
$\overline\Omega$ data demand that we explore without prejudice
techniques for measuring kaon fluctuations.

A further implication of Ref.~\cite{KapustaWong} is that the DCC
regions must be rather small, with a size of about $2$~fm. Such a size
is consistent with predictions based on dynamical simulations of the
two flavor linear sigma model \cite{GGP1}. More importantly, the DCC
search for anomalous event structure in neutral and charged pions by
the WA98 Collaboration at the SPS has revealed no evidence of large
DCC "domains" \cite{qmrev2}. Note that I loosely use the word "domain"
to refer to spatial regions in which the condensate is somehow
coherent. I stress that no thermodynamically stable domain structures
are expected in theoretical descriptions of DCC.

In this paper I study kaon isospin fluctuations in the presence of many
small DCC. In the next section I discuss how DCC may lead to kaon 
fluctuations. Following Ref.~\cite{GavinKapusta}, I 
compute the probability distribution that describe the DCC contribution
to these fluctuations and combine the DCC fluctuations
with a contribution from a random thermal background. 
Pion fluctuations due to many small DCC have
been addressed by Amado and Lu \cite{AmadoLu} and Chow and Cohen
\cite{ChowCohen}. Here we focus 
on kaon fluctuations; pion as well as kaon results are presented in
Ref.~\cite{GavinKapusta}. In sec.~4 I
assess robust statistical observables that can be used to measure the
impact of many small DCC at RHIC and LHC. In sec.~5 I present work in
progress on event generator simulations to understand the magnitude
and centrality dependence of the statistical observables in the
absence of DCC. In particular, I obtain a dynamic isospin
fluctuation observable analogous to the dynamic charge observable used
to measure net charge fluctuations at RHIC
\cite{Voloshin}. Of the quantities considered, this observable
isolates the DCC effect from other sources of fluctuations best.

\section{Strange DCC}

To illustrate how a strange DCC can form, let us first consider QCD with
only up and down quark flavors. Equilibrium high temperature QCD
respects chiral symmetry if the quarks are taken to be massless. This
symmetry is broken below $T_c\sim 150$~MeV by the formation of a
chiral condensate $\langle \sigma\rangle \sim \langle
\overline{u}u+\overline{d}d\rangle$ that is a scalar isopin singlet. 
However, chiral symmetry implies that $\sigma$ is degenerate with a
pseudoscalar isospin triplet of fields with the same quantum numbers
as the pions. In reality, chiral symmetry is only approximate and the
140~MeV pion mass is different from the $800\pm 400$~MeV mass of the
leading sigma candidate \cite{pdg}. Nevertheless, lattice calculations
exhibit a dramatic drop of $\langle \sigma\rangle$ near $T_c$ at
finite quark masses.

A DCC can form when a heavy ion collision produces a high energy
density quark-gluon system that then rapidly expands and cools through
the critical temperature. Such a system can initially break chiral
symmetry along one of the pion directions, but must then evolve to the
$T=0$ vacuum by radiating pions. A single coherent DCC radiates a
fraction $f_\pi$ of neutral pions compared to the total that satisfies
the probability distribution
\begin{equation}
\rho_1(f_\pi) = {{1}\over{2f_\pi^{1/2}}}\,\,\,\,\,\,\,\,\,\,\,\,\,\,\,\, 0 < 
f_\pi \le 1,
\end{equation}
\cite{Anselm,Blaizot,Bjorken}. Such isospin fluctuations constitute the 
primary signal for DCC formation in the pion sector. The enhancement
of baryon-antibaryon pair production is a secondary effect due to the
relation between baryon number and the topology of the pion condensate
field \cite{KapustaSrivastava}.

This two flavor idealization only applies if the strange quark mass
$m_s$ can be taken to be infinite. Alternatively, if I take $m_s =
m_u = m_d =0$, then the chiral condensate would be an up-down-strange
symmetric scalar field. The more realistic case of $m_s\sim 100$~MeV
is between these extremes, so that $\langle \sigma\rangle \sim \langle
\cos\theta(\overline{u}u+\overline{d}d) +
\sin\theta(\overline{s}s)\rangle$. The mixing angle $\theta$ is highly
uncertain since it depends on the sigma mass together with the $\pi,
K, \eta$ and $\eta^\prime$ masses and the $\eta-
\eta^\prime$ mixing angle \cite{GGP2}. A disoriented condensate can evolve by 
radiating $\pi, K, \eta$ and $\eta^\prime$ mesons, with the neutral pion 
fraction satisfying (1). Randrup and Sch\"affner-Bielich find that the kaon 
fluctuations from a single large DCC satisfy \cite{Randrup}
\begin{equation}
\rho_1(f_K) = 1\,\,\,\,\,\,\,\,\,\,\,\,\,\,\,\, 0 \le f_K \le 1,
\end{equation}
where $f_K = (K^0 + \overline{K}^0)/(K^+ + K^- + K^0 +
\overline{K}^0)$. Moreover, the condensate fluctuations can now produce
strange baryon pairs \cite{KapustaWong}. Linear sigma model
simulations indicate that pion fluctuations dominate three-flavor DCC
behavior, while the fraction of energy imparted to kaon fluctuations
is very small due to the kaons' larger mass. On the other hand, domain
formation may be induced by other mechanisms such as kaon condensation
at high baryon density \cite{Kaplan}, bubble formation \cite{qm95} or
decay of the Polyakov loop condensate
\cite{Pisarski}. 

\section{DCC Mesons from Many Small Domains}

Why does the DCC's size matter? Pion measurements in individual
collision events can distinguish DCC isospin fluctuations from a
thermal background only if the disoriented region is sufficiently
large \cite{qmrev1}. DCC can then be the dominant source of pions at
low transverse momenta, since $\langle p_t \rangle\sim 1/R$ for a
coherent region of size $R$. Experiments focusing on low $p_t$ can
study neutral and charged pion fluctuations \cite{Bjorken}, wavelet
\cite{Ina} and HBT signals \cite{qmrev1,HBT} to extract detailed information. 
In contrast, for small domains ($R<3$~fm \cite{qmrev1}) DCC signals
are hidden by fluctuations due to ordinary incoherent production
mechanisms. This holds even if many such regions are produced per
event. DCC mesons from small regions may have momenta of a few hundred
MeV, nearer the $pp$ mean value. Different regions would not add
coherently to alter HBT, nor would their small spatial structures
affect wavelet analysis.

Importantly, baryon pair enhancement \cite{KapustaWong} is substantial
only if there are many small incoherent regions. The large winding
numbers that produce baryon-antibaryon pairs require many small
regions with random relative orientations of the pion field. To
describe strange antibaryon enhancement, Kapusta and Wong assume
roughly 100 DCC regions of size roughly $2$~fm
\cite{KapustaWong}. Topological models of baryon-antibaryon pair
production successfully describe $e+e-$ and hadronic collision data
\cite{EllisKowalski}. The connection of DCC to topological pair
production was pointed out in Ref.~\cite{KapustaSrivastava}; see also
\cite{DeGrand}.

To compute the distribution of kaons due to many small DCC regions,
define $f = (K^0 + \overline{K}^0)/(K^+ + K^- + K^0 +
\overline{K}^0)$.  To an excellent approximation the number of neutral
kaons is equal to twice the number of short-lived neutral kaons $K_S$
which are more readily measurable in high energy heavy ion collisions.
The fraction $f$ ranges from 0 to 1.  The statistical distribution in
$f$ for a single domain is $\rho_1(f) = 1$.  The distribution for $n$
randomly oriented, independent domains is
\begin{equation}
\rho_n(f) = \int  \prod_{k=1}^n df_k \, 
\rho_1(f_k) \, \delta\left( f - \frac{1}{n} \sum_{j=1}^n f_j \right) \, .
\end{equation}
In \cite{GavinKapusta} I obtain
\begin{equation}
\rho_n(f) = n^2 \sum_{0 \leq k < n(1-f)} (-1)^k
\frac{[n(1-f)-k]^{n-1}}{k! (n-k)!}\, .
\end{equation}
In the limit that $n \gg 1$, this distribution tends toward a Gaussian
of mean $\langle f\rangle =1/2$ and standard deviation $\sigma =
\{12n\}^{-1/2}$. Results for pions are presented in
\cite{GavinKapusta}.

In a more realistic scenario some kaons will come from the decay or
realignment of DCC domains and some will come from more conventional
sources.  I shall refer to the latter as random or thermal, even
though that may be a bit of a misnomer.  What I mean by random or
thermal is that the distribution of kaons from non-DCC sources is
\begin{equation}
\rho_0(f_0) = \frac{1}{2\pi \sigma_0^2} \exp \left[ -(f_0-1/2)^2
/2\sigma_0^2 \right] \, .
\end{equation}
For a completely random source the width $\sigma_0$ is related to the total 
number $N_{\rm random}$ of non-DCC kaons by
\begin{equation}
\sigma^2_0 = \frac{1/2(1-1/2)}{N_{\rm random}} = \frac{1}{4N_{\rm random}} \, .
\end{equation}

Now let us assume that a fraction $\alpha_K$ of all kaons come from non-DCC 
sources and the remaining fraction $\beta_K = 1-\alpha_K$ come from $n \gg 1$ 
independent DCC domains.  Letting $N$ denote the total number of 
kaons, I have $N_{\rm random} = \alpha_K N$ and $N_{\rm DCC} = \beta_K N$.  
Folding together two Gaussians gives a Gaussian.
\begin{eqnarray}
\rho_K(f) &=& \int df_0 df_n \rho_0(f_0) \rho_n(f_n) \delta (f - \alpha_K f_0
-\beta_K f_n) \nonumber \\
&=& \frac{1}{\sqrt{2\pi \Delta_K^2}} \exp \left[ -(f-1/2)^2/2\Delta_K^2 \right]
\end{eqnarray}
For a thermal source plus $n$ DCC domains, the net width is
\begin{equation}
\Delta_K^2 = \frac{\alpha_K}{4N} + \frac{\beta_K^2}{12n} =
\frac{1}{4N} + \left\{ \frac{\beta_K^2}{12n} - \frac{\beta_K}{4N}
\right\}\, .
\end{equation}
The expression in curly brackets at the end represents the difference between 
the actual width and the width the distribution would have if there was no 
contribution from DCC kaons.  This change in the width may be positive or 
negative, depending on the parameters.

\section{Statistical Analysis}

Detection of small incoherent DCC regions in high energy heavy ion collisions 
requires a statistical analysis in the $\pi^0\pi^\pm$ or the $K_S^0K^\pm$ 
channels. Neutral mesons can be detected by the decays $\pi^0\rightarrow 
\gamma\gamma$ or $K^0_S\rightarrow \pi^+\pi^-$. The analysis we propose in
\cite{GavinKapusta} is sensitive to correlations due to isospin fluctuations.
We expect these correlations to vary when DCC regions increase in
abundance or size as centrality, ion-mass number $A$, or beam energy
are changed. Correlation results combined with other signals, such as
baryon enhancement \cite{KapustaWong}, can be used to build a
circumstantial case for DCC production.

Correlations of $\pi^0\pi^\pm$ and $ K_S^0 K^\pm$ can be determined by
measuring the robust isospin covariance,
\begin{equation}
R_{c0} = {{\langle N_cN_0\rangle - \langle N_c\rangle\langle N_0\rangle}
\over{\langle N_c\rangle\langle N_0\rangle}},
\end{equation}
where $N_0$ and $N_c$ are the number of neutral and charged mesons. I
take $N_0= N_{\pi^0}$ and $N_c=N_{\pi^+}+N_{\pi^-}$ for pion
fluctuations and $N_0=2N_{ K^0_S}$ and $N_c=N_{K^+}+N_{K^-}$ for kaon
fluctuations. The ratio (9) has two features that are convenient for
experimental determination. First, this observable is independent of
detection efficiency as are the ``robust'' ratios discussed in
\cite{Taylor}. Robust observables are useful for DCC studies because
charged and neutral particles are identified using very different
techniques and, consequently, are detected with different
efficiency. Observe that robust quantities are not affected by the
unobserved $K_L^0$, since the strong-interaction eigenstates $K^0$ and
${\overline K}^0$ are a superposition $K_L^0$ and $K_S^0$ until their
decay well outside the collision region.  Second, since (9) is
obtained from a statistical analysis, individual $\pi^0\rightarrow
\gamma\gamma$ or $K^0_S\rightarrow \pi^+\pi^-$ need not be fully
reconstructed in each event. This feature is crucial because it would
be extraordinarily difficult -- if not impossible -- to reconstruct a
low momentum $\pi^0$ in heavy ion collisions except on a statistical
basis.

Next I define robust variance
\begin{equation}
R_{aa} = {{\langle N_a^2\rangle - \langle N_a\rangle^2 - \langle N_a\rangle}
\over{\langle N_a\rangle^2}},
\end{equation}
where $a = c$ or 0. To see why (10) is robust, denote the
probability of detecting each meson $\epsilon$ and the probability
of missing it $1-\epsilon$. For a binomial distribution the average number of 
measured particles is $\langle N_a\rangle^{\rm exp} = \epsilon\langle 
N_a\rangle$ while the average square is $\langle N_a^2\rangle^{\rm exp} = 
\epsilon^2\langle
N_a^2\rangle+\epsilon(1-\epsilon)\langle N_a\rangle$. I then find
\begin{equation}
R_{aa}^{\rm exp} = R_{aa},
\end{equation}
independent of $\epsilon$ \cite{Pruneau}; the proof that (9) is
robust is similar. The ratios (9) and (10) are strictly robust only
if the efficiency $\epsilon$ is independent of multiplicity. Further
properties and advantages of these and similar quantities are
discussed in \cite{Pruneau}.

To study DCC fluctuations I define the dynamic isospin observable
\begin{equation}
\nu_{\rm dyn}^{c0} = R_{cc} + R_{00} - 2R_{c0}.
\end{equation}
Analogous observables have been employed to study net charge
fluctuations in particle physics \cite{Whitmore,Boggild} and were
considered in a heavy ion context in \cite{Voloshin} and
\cite{Mrowczynski}. This quantity can be written in terms of
\begin{equation}
\nu^{c0} = \left\langle\left({{N_0}\over{\langle N_0\rangle}} -
{{N_c}\over{\langle N_c\rangle}} \right)^2\right\rangle.
\end{equation}  
To isolate the dynamical isospin fluctuations from other sources
of fluctuations, one obtains (12) by subtracting from (13) the
uncorrelated Poisson limit $\nu_{\rm stat}^{c0}=\langle
N_0\rangle^{-1}+\langle N_c\rangle^{-1}$.  Indeed, we show in (19)
below that the quantity (12) depends primarily on the fluctuations of
the neutral fraction $f$, while the individual ratios (9) and (10)
have additional contributions.

I illustrate the effect of DCC on the dynamic isospin fluctuations by
writing $N_0 = fN$ and $N_c = (1-f)N$.  Small fluctuations on $f$ or $N$
results in the changes
\begin{equation}
\frac{\Delta N_0}{\langle N_0 \rangle} = {{\Delta N}\over{\langle N\rangle}}
+ {{\Delta f}\over{\langle f\rangle}}, \,\,\,\,\,\,\,\,\, \,\,\,\,\,\,\,\,\, 
\frac{\Delta N_c}{\langle N_c \rangle} = {{\Delta N}\over{\langle N\rangle}}
 -  {{\Delta f}\over{1-\langle 
f\rangle}} \, . 
\end{equation}
I obtain the average
\begin{equation}
{{\langle\Delta N_0^2\rangle}\over{\langle N_0\rangle^2}}
 = v + {{2c}\over{\langle N\rangle \langle
f\rangle}} + {{\Delta^2}\over{\langle f\rangle^2}}. 
\end{equation}
Here the contribution of the variance of the total number of mesons
is $v \equiv \langle \Delta N^2\rangle/\langle N\rangle^2$ and the
charge-total covariance is $c \equiv \langle \Delta N\Delta f\rangle$.
DCC formation primarily effects the charge fluctuation contribution,
$\Delta^2 \equiv \langle (\Delta f)^2\rangle$, from (15) or (17). Similarly,
\begin{equation}
{{\langle\Delta N_c^2\rangle}\over{\langle N_c\rangle^2}} 
= v - {{2c}\over{\langle N\rangle(1-\langle
f\rangle)}} + {{\Delta^2}\over{(1-\langle f\rangle)^2}} \, ,
\end{equation}
and
\begin{equation}
%{{\langle\Delta N_c\Delta N_0\rangle}\over{\langle N_c\rangle\langle 
%N_0\rangle}} 
R_{c0}= v + \left({{1}\over{\langle f\rangle}} - {{1}\over{1-\langle 
f\rangle}}\right){{c}\over{\langle N\rangle}}
- {{\Delta^2}\over{(1-\langle f\rangle)^2}} \, 
\end{equation}
where $R_{c0}$ is given by (9). Using (21) I get 
%\begin{equation}
%\nu_{\rm dyn}^{c0} = {{\Delta^2}\over{[\langle f\rangle(1-\langle 
%f\rangle)]^2}}
%-  {{1}\over{\langle f\rangle(1-\langle f\rangle)\langle N\rangle}} \, .
%\end{equation}
\begin{equation}
\nu_{\rm dyn}^{c0} = {{1}\over{\langle f\rangle(1-\langle f\rangle)}}
\left({{\Delta^2}\over{\langle f\rangle(1-\langle 
f\rangle})} -{{1}\over{\langle N\rangle}}\right).
\end{equation}
This observable isolates the isospin fluctuations, whereas the
individual $R_{ab}$ depend on the fluctuations in total meson number,
$v$ and $c$ as well.

I estimate the effect of DCC on the dynamical fluctuations (18) using
(6) and (8). I take $\langle N\rangle = N_K$ for kaons and $\langle
N\rangle = N_{\pi}$ for pions; these are the total number of mesons of
the indicated kind.  For kaons
\begin{equation}
\nu_{\rm dyn}^{c0}({\rm K~DCC}) = 4\beta_K \left( \frac{\beta_K}{3n}
- \frac{1}{N_K} \right) \, .
\end{equation}
These quantities can be positive or negative depending on the
magnitude of $\beta$ compared to the number of domains per kaon. In
fact the dynamical fluctuation may even be positive for one kind of
meson and negative for the other. 

\section{Fluctuations in the Absence of DCC}

Let us now discuss work in progress in which Abdel-Aziz and I use
event generators to simulate conventional sources of kaon fluctuations
\cite{hamlet}. In the absence of DCC, $\alpha = 1$ and $\beta = 0$ so
that (19) implies $\nu_{\rm dyn}^{c0} \equiv 0$. On the other hand,
incomplete equilibration may result in dynamical correlations in
nuclear collisions not described by (5). Little is known from $pp$
experiments about kaon fluctuations. Event generators such as HIJING
and URQMD models both yield negative values of $\nu_{\rm dyn}^{c0}$ in pp
collisions. HIJING simulations of central Au+Au at 200 $A$ GeV in the
rapidity range $-0.5 < y < 0.5$ yield $\nu_{\rm dyn}^{c0} \approx
-0.002$ for 47 $K^+$ and 44 $K_S^0$ on average
\cite{HIJING}. 

The onset of DCC formation can substantially change the value of
$\nu_{\rm dyn}^{c0}$. To search for this onset it is useful to study
fluctuations as a function of collision centrality. I use HIJING to
estimate the influence of conventional collision geometry and dynamics
on the centrality dependence. However, I find that one can understand
the HIJING results quite simply using the wounded nucleon model. In a
multiple collision models such as the wounded nucleon model, one
describes a nucleus-nucleus collision as a superposition of M
independent nucleon-nucleon sub-collisions. The robust
variance and covariance satisfy
\begin{equation}
R_{ab}={{r_{ab}}\over{\langle M\rangle}}+{{\langle M^2\rangle-\langle
M\rangle^2}\over{\langle M\rangle^2}}\, ,
\end{equation} 
where the $r_{ab}$ are coefficients describing the fluctuations in
each sub-collision and $a,b$ can equal either $c$ or $0$. The
dynamical isospin observable,
\begin{equation}
\nu_{\rm dyn}^{c0}
=R_{cc}+R_{00}-2R_{c0}={{\nu_{0}}\over{\langle M\rangle}},
\end{equation}
is independent of the contribution from the fluctuations of $M$. In
the wounded nucleon model, $M$ is given at each impact parameter by
the number of participant nucleons.

\begin{figure}
\begin{center}
\includegraphics[width=3in]{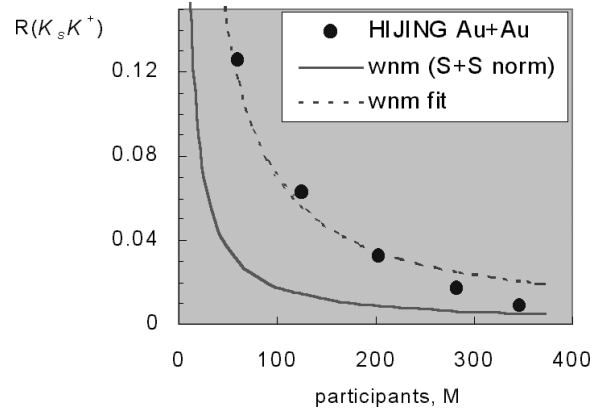}
\end{center}
\caption{
Scaled covariance for HIJING compared to the wounded
nucleon model normalized to HIJING S+S (solid curve) and fit (dashed).}
\label{fig1}
\end{figure}

\begin{figure}
\begin{center}
\includegraphics[width=3in]{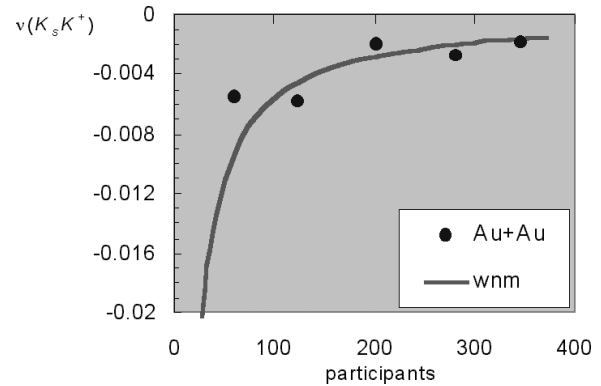}
\end{center}
\caption{Dynamic isospin observable for the 
same simulated collisions. Wounded nucleon result agrees with 
HIJING S+S normalization. }
\label{fig2}
\end{figure}
In figs.~1 and 2, I show the kaon covariance (9) and dynamical isospin
observable (12) as functions of the number of participants $M$
computed from 50,000 HIJING events for Au+Au at 200 A$GeV$ in the
rapidity range $-0.5<y<0.5$ compared to wounded nucleon model
calculations (20). The number of participants at impact parameter $b$
for a symmetric Au+Au collision is computed using $M(b)=2\int ds
T(s)\{1-e^{-\sigma_{NN}T(b-s)}\}$, where $T(b) = \int
\rho(z,b)dz$ is the familiar nuclear thickness function and $\rho$ is
the three-parameter Fermi nuclear density for Au. In fig.~1, the solid
curve is determined using (20) with a coefficient $r_{c0}$ computed
from 50,000 S+S collisions, while the dashed curve is obtained by
varying $r_{c0}$ to fit Au+Au. The HIJING results scale roughly as
$R_{c0}\propto M^{-1}$ as expected, but the difference from the
wounded nucleon model are rather large.

Figure 2 shows the dynamical isospin observable as a function of $M$
from HIJING. The solid curve is obtained from (21) with $\nu_0$
determined from HIJING S+S. I find that the wounded nucleon model is
in excellent agreement with HIJING, suggesting that the correlations
that increase $R_{ab}$ compared to the wounded nucleon model are
similar for all kaon charge states. The agreement of HIJING and the
wounded nucleon model in fig.~2 is likely due to the
following. First, baryon stopping is unimportant at the high RHIC
energy. Different numbers of protons and neutrons would alter the
isospin balance. Second, high $p_T$ aside, the way HIJING describes soft
interactions is quite similar to the wounded nucleon model, since it
doesn't incorporate final state cascading. To understand the effect of
cascading, we are currently studying URQMD collisions \cite{hamlet}.

\begin{figure}[htb]
\centerline{\includegraphics[width=3in]{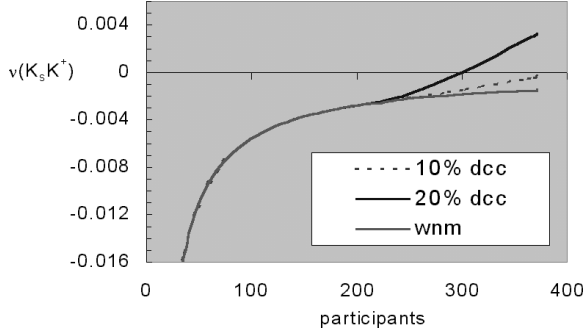}}
\caption[]{Scaled kaon covariance vs. number of participants as above plus a
contribution from 10 DCC domains.}
\label{fig3}
\end{figure}
To illustrate the possible scenario for the onset of DCC effects, I
assume that DCC kaons add to the kaons from multiple sub-collisions
according to
\begin{equation}
\nu_{\rm dyn}^{c0} = \beta^2\nu_{dcc}+ (1-\beta^2)\nu_{wnm},
\end{equation}   
where $\beta$ is the fraction of DCC kaons, $\nu_{dcc}$ is the DCC
contribution given by (10) and $\nu_{wnm}$ is given by (21). I assume
that DCC production above an impact parameter $b_0$ exhibits a
threshold behavior, $\beta = \beta_0[1-(b/b_0)^2]$, where $b_0$ and
$\beta_0$ are ad hoc constants. In fig.~4, I show estimates assuming
that 10 domains contribute kaons in the range $-0.5 < y < 0.5$ for
$b_0 \sim 6$~fm, varying the dcc fraction $\beta_0$ between 10\% and
20\%. Note that the DCC contribution to $\nu$ is positive for these
values.

\section{Discussion and Conclusion}

Reference \cite{KapustaWong} argued that the anomalous abundance and
transverse momentum distributions of $\Omega$ and $\overline{\Omega}$
baryons in central collisions between Pb nuclei at 17 $A$ GeV at the
CERN SPS is evidence that they are produced as topological defects
arising from the formation of many domains of disoriented chiral
condensates (DCC) with an average domain size of about 2 fm. Motivated
by this interpretation, Kapusta and I studied the effect of DCC on the
distribution of the fractions of neutral kaons and pions in
\cite{GavinKapusta}. 

The DCC pioneers \cite{Anselm,Blaizot,Bjorken,DCCreview} had hoped
that a large percentage of pions might be emitted from just a few big
domains, on the order of 5 to 8 fm (kaons were not considered). Such
large domains have been ruled out at SPS \cite{qmrev2}, but remain
possible at RHIC. More conservatively, as the number of domains grow
and their average size diminishes, the impression left on the
fluctuations in the neutral fraction becomes more subtle and less
unique.  For many small domains, statistical measurements of both
neutral kaons (pions) and charged kaons (pions) are needed to observe
the rather small isospin fluctuations. We have identified robust
observables for that purpose.  In particular, we have shown that the
dynamical isospin observable (12) is sensitive to DCC but not to
thermal or multiple-collision sources of fluctuations, as discussed in
the text near eqs.\ (19) and (21) respectively. While HIJING
simulations support this conclusion, more work remains. For example,
one can use URQMD to study the effect of final state scattering on
this observable. Isospin fluctuations can appear as changes in the
magnitude of the dynamical isospin observable as centrality is
varied. We emphasize that similar consequence may follow from any
mechanism that produces many small domains that decay to pions and
kaons, such as the Polyakov Loop Condensate \cite{Pisarski}.

%\section*{Acknowledgements}

I am grateful to Joe Kapusta for a very enjoyable collaboration and
thank M. Abdel-Aziz, R. Bellwied, C. Pruneau, S. Voloshin and,
especially, H.\ St\"ocker for many useful discussions. This work was
supported in part by U.S. DOE Grant number DE-FG02-92ER40713.

\end{document}